\documentclass{aastex61}

\received{mmm dd, 2017}
\revised{mmm dd, 2017}
\accepted{mmm dd, 2017}
\submitjournal{ApJ}

\shorttitle{The Magnetic Field in Solar Filaments}
\shortauthors{Y. Hanaoka \& T. Sakurai}

\begin{document}

\title{Statistical Study of the Magnetic Field Orientation in Solar Filaments}

\correspondingauthor{Yoichiro Hanaoka}
\email{hanaoka@solar.mtk.nao.ac.jp}

\author{Yoichiro Hanaoka}
\affil{National Astronomical Observatory of Japan \\
2-21-1 Osawa, Mitaka \\
Tokyo, 181-8588, Japan}

\author{Takashi Sakurai}
\affil{National Astronomical Observatory of Japan \\
2-21-1 Osawa, Mitaka \\
Tokyo, 181-8588, Japan}

\begin{abstract}
We have carried out a statistical study of the average orientation of the magnetic field in
solar filaments with respect to their axes for more than 400 samples, based on data taken with
daily full-Sun, full-Stokes spectropolarimetric observations using
the \ion{He}{1} 1083.0 nm line. The major part of the samples are the filaments in the quiet areas, but those in the active areas are included as well.  
The average orientation of the magnetic field in filaments shows a systematic property depending on the hemisphere;
the direction of the magnetic field in filaments in the northern (southern) hemisphere mostly deviates clockwise
(counterclockwise) from their axes, which run along the magnetic polarity inversion line. 
The deviation angles of the magnetic field from the axes, are concentrated between 10--30$^\circ$.
This hemispheric pattern is consistent with that revealed for chirality of filament barbs, filament channels and for other solar features found to possess chirality. 
For some filaments it was confirmed that their magnetic field direction is locally parallel to their structure seen in H$\alpha$ images.
Our results for the first time confirmed this hemispheric pattern with the direct observation of the magnetic field in filaments. 
Interestingly, the filaments which show the opposite magnetic field deviation to the hemispheric pattern, are in many cases found above the polarity inversion line whose ambient photospheric magnetic field has the polarity alignment being opposite to that of active regions following the Hale-Nicholson law.
\end{abstract}


\section{Introduction} \label{sec:intro}

The magnetic field in solar filaments and in their environment govern their formation and evolution, as well as their eruption to the interplanetary space. Despite its importance, limited measurements have so far been carried out, because of technical obstacles.  The most useful spectral line is the \ion{He}{1} 1083.0 nm line to measure the magnetic field of the cool plasma in the filament, which is formed in chromospheric conditions.  In recent years,
a method to interpret the polarization of the \ion{He}{1} 1083.0 nm line
taking the atomic polarization into account was established
\citep{TrujilloBueno2002}, and some detailed case studies of the magnetic
field in filaments have been done \citep[e.g.,][]{Kuckein2009,
OrozcoSuarez2014}. However, there are only a few statistical studies.
\citet{Bommier1998} reviewed a significant number of the
measurements of the average magnetic field in prominences \citep{Leroy1983, Leroy1984, Bommier1986, Bommier1994}, and concluded that the
majority of prominences have magnetic fields directed from the negative polarity
side on the photosphere to the positive polarity side with certain angles between the field vector and the neutral line, that generally depend on the hemisphere. 

On the other hand, there are a number of indirect studies of the characteristics of the magnetic field in filaments.
\citet{Zirker1997} and \citet{Martin1998} reviewed the fine structures in filaments seen in H$\alpha$ \citep[studied by e.g.,][]{Martin1994}, which
are considered to reflect the magnetic field in filaments, and concluded that
 the filament
fine structures, particularly barbs of filaments, show a chiral nature that 
in the northern hemisphere, barbs are right-bearing (dextral),
and in the southern hemisphere, they are left-bearing (sinistral).  
This property of the fine structure in filaments has been further studied and discussed by  \citet{Pevtsov2003}, \citet{Bernasconi2005}, \citet{Yeates2007}, and \citet{Ouyang2017}.

The hemispheric pattern of the chirality of the magnetic field in filaments is one of the manifestations of the chiral nature of the global solar dynamo action in the northern and the southern hemispheres.  There are various features showing the chirality depending on the hemisphere as well as the filament magnetic field; the current helicity in active regions \citep[e.g.,][]{Pevtsov1995, Hagino2005}, the structure seen in filament channels \citep{Foukal1971, Martin1994, Sheeley2013}, the whorl structure around sunspots \citep[e.g.,][]{Hale1927, Richardson1941}, the skew of coronal loops \citep[][]{Rust1996,Martin1996, Lim2009}, and interplanetary clouds associated with coronal mass ejections (CMEs) \citep{Rust1994, RustKumar1994, Bothmer1994}.
Particularly, the magnetic field in filament channels, filaments, and coronal loops above filaments are considered to be a part of the unified magnetic field structure, and it is also known that such structures are often related to eruptive solar events that
produce interplanetary clouds.
In this paper, we focus on the magnetic field structure in filaments, which is one of the keys to understand the dynamo action of the Sun and the eruptive events on the solar surface.

A full-disk, full-Stokes spectropolarimeter to take synoptic magnetic field
maps of the Sun in near infrared wavelengths, which was installed on the Solar Flare
Telescope \citep{Sakurai1995}, has been in operation since 2010 at the
Mitaka headquarters of the National Astronomical Observatory of Japan
\citep[Sakurai et al. in preparation; see also][]{Hanaoka2011}. One of its
observing wavelength bands includes the \ion{He}{1} 1083.0 nm line and the photospheric \ion{Si}{1} 1082.7 nm
line, and the other includes the \ion{Fe}{1} 1564.8 nm and 1565.3 nm line pair.
Polarization maps taken in the daily observation show polarization signals
of filaments \citep{Hanaoka2014}. Therefore, polarization data of quite a lot
of filaments have been accumulated. Linear polarization
signals of the \ion{He}{1} 1083.0 nm line in filaments reflect the magnetic field of 
filaments. In this paper, we report results of a statistical
investigation of the polarization in filaments, concentrating on the
orientation of the linear polarization of the \ion{He}{1} 1083.0 nm line with respect to the axes of the filaments.  This is a study based on the direct observation of the magnetic field of filaments.

\section{Observation and Sample Selection} \label{sec:obs}

\begin{figure}
\epsscale{1.16}
\plotone{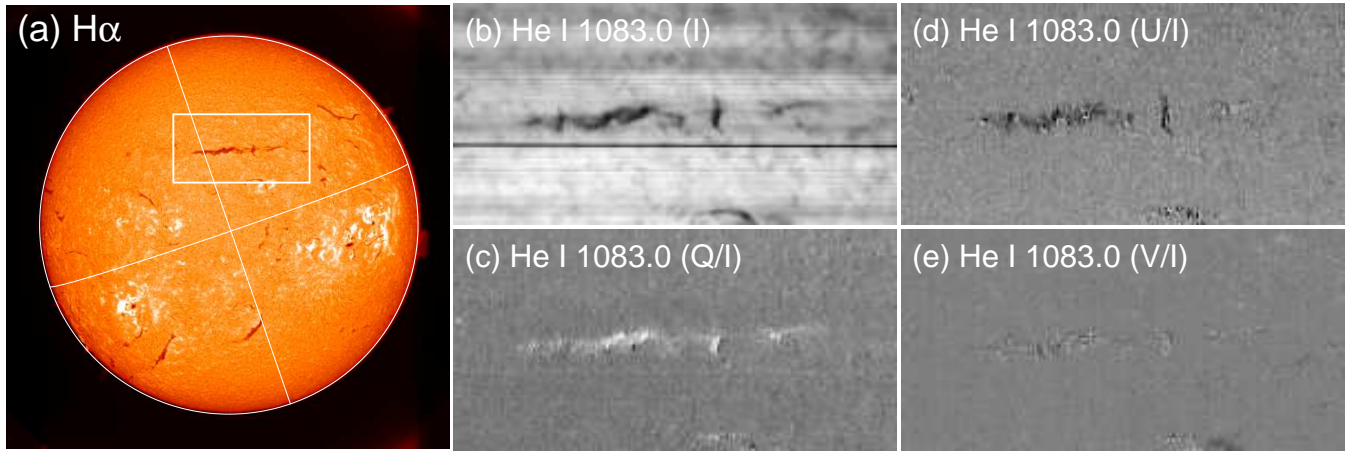}
\caption{
H$\alpha$ and polarization data of a filament on 2014 November 23.  (a) Full-disk H$\alpha$ image.  A box ($11'.7\times5'.8$) indicates the filament.  (b)--(e) Stokes maps of the box in panel (a) in the \ion{He}{1} 1083.0 nm line.  A horizontal dark line in panel (b) is due to a particle of dust on the slit.  In panels (c)--(e), the range of $\pm 0.5$ \% of the degree of polarization is shown, and the celestial east-west direction is defined as the positive-$Q$ axis.
The celestial north is to the top in all the panels.
\label{fig:fig1}}
\end{figure}

\begin{figure}
\plotone{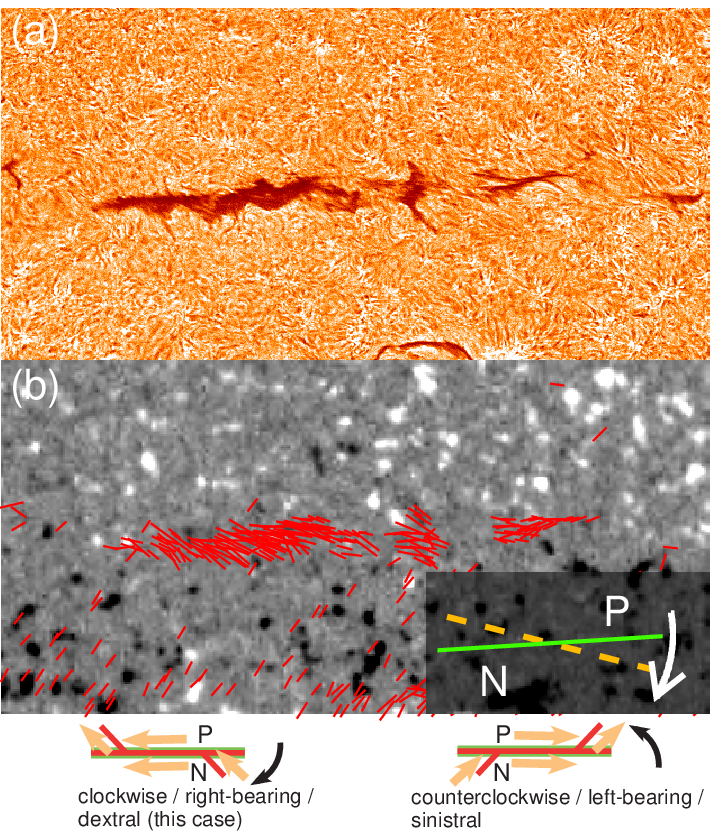}
\caption{
(a) Enlarged H$\alpha$ image of the filament shown in Figure \ref{fig:fig1}.  The contrast is enhanced to show the structures in the filament clearly.  (b) The linear polarization signals averaged over $2\times2$ pixels for the degree of polarization $>$ 0.1 \%  represented by red lines on the gray scale Stokes $V/I$ image of the \ion{Si}{1} 1082.7 nm line.  White and black correspond to the positive and negative polarities.
In the inset at the lower-right corner of panel (b),  the approximate direction of the axis of the filament is shown with a solid green line, and the average direction of the linear polarization signals is shown with a dashed orange line.  Labels P and N indicate the positive and negative polarity sides facing each other across the filament axis.  In the schematic illustrations at the bottom, the spine and barb  structure is shown with red lines along with the filament axis shown with green lines.  
The magnetic field in the spine and barb  structure, which is assumed to direct from the negative side to positive side, are shown by orange arrows.
Both the 
``right-bearing'' (the case of the filament in this Figure) and ``left-bearing'' chirality cases are illustrated.  The deviation of the direction of the polarization in filaments from their axes are also shown, on the assumption that the polarization is parallel to the H$\alpha$ barbs.
\label{fig:fig2}}
\end{figure}

\subsection{Polarization signals in filaments}

Figure \ref{fig:fig1} shows an example of the polarization data taken with the spectropolarimeter of the Solar Flare Telescope.  Panel (a) shows a full-disk H$\alpha$ image on 2014 November 23 taken with the filtergraph also installed on the Solar Flare Telescope.  There is an ordinary quiescent filament in the box ($11'.7\times5'.8$) in panel (a).
Stokes parameter maps of
the \ion{He}{1} 1083.0 red component for the box in panel (a) are shown in panels (b)--(e).  For the details of the \ion{He}{1} 1083.0 multiplet and its polarization, see \citet{TrujilloBueno2002} and \citet{TrujilloBueno2007}.
The Stokes $I$, $Q/I$, and $U/I$ maps (panels (b)--(d)) show
the integrated polarization data in a 0.4 \AA\ range around the center of
the \ion{He}{1} 1083.0 red component, while the Stokes $V/I$ map in panel (e) shows polarization data of which spectral profiles are anti-symmetric with respect to the line center wavelength (namely, the
Zeeman polarization component).  The original polarization data have a pixel size of about $1''.75\times1''.75$, but actual spatial resolution is typically 3--5$''$ due to the blurring by the seeing effect.  
Therefore, each pixel in these Stokes maps was resized to $3''.5\times3''.5$.  The root-mean-square noise level of the polarization data is about $3\times10^{-4}$ in the degree of polarization.

The filament, which is dark in the Stokes $I$ map (panel (b)) as in the H$\alpha$ image,  shows conspicuous polarizations in the Stokes $Q/I$ and $U/I$
maps  (panels (c) and (d)), but the $V/I$ map (panel (e)) only shows noise at the site of the filament.  The polarizations in the filament seen in panels (c) and (d) are net linear polarizations. These polarizations are produced by the atomic level polarization, and the direction
of the linear polarization of the \ion{He}{1} 1083.0 red component is parallel to the magnetic field
\citep{TrujilloBueno2002}. Therefore, the direction
of the linear polarization shows the direction of the magnetic field in the filament material,
leaving the 180$^\circ$ azimuth ambiguity unresolved. The degree of the linear polarization in the filament in Figure
\ref{fig:fig1} is typically of the order of 0.1 \%, and the corresponding magnetic field
strength is roughly several gauss \citep{TrujilloBueno2002}.

The structures seen in the H$\alpha$ image and the polarization signals of the \ion{He}{1} 1083.0 nm line in the
filament are shown in Figure \ref{fig:fig2} with enlarged views.
In panel (b), linear polarization signals of the
pixels whose degree of polarization exceeds 0.1 \% are
shown with red lines. The direction and length of each red line
correspond to the direction and the degree of polarization. 
This panel contains $200\times100$ pixels, but the polarization signals of every $2\times2$ pixels are displayed, not to make the figure too busy.
The noise of about $3\times10^{-4}$ included in the polarization data of the order of 0.1 \% does not produce significant errors in the direction of the polarization; the error is typically several degrees or less.  
The background gray-scale image in panel (b) is a Stokes $V/I$ image of the \ion{Si}{1} 1082.7 nm line showing the photospheric magnetic field.  It can be confirmed that the filament is located above the polarity inversion line seen in this background image.  The \ion{Si}{1} 1082.7 data were simultaneously taken with the \ion{He}{1} 1083.0  data, and therefore, their quality is equivalent to that of the \ion{He}{1} 1083.0 data.  Furthermore, 
there is no ambiguity in the co-alignment between the \ion{He}{1} 1083.0 image and the \ion{Si}{1} 1082.7 image.  

The H$\alpha$ filament in Figure \ref{fig:fig2}(a) is a typical one in which the spine and barbs are viewed nearly from the top down. The spine is considered to follow the polarity inversion line in panel (b).  However, the spine is not very clear, because from this view of a filament, the spine is exceedingly narrow,
and it has the common problem of being masked by the barbs on the equatorward side of the filament.
On the other hand, the filament seems to consist of thread-like structures, which generally run from the upper-left to the lower-right.  These are mainly barbs, which branch off from the spine.  Previously published high-resolution images show that filaments consist of numerous very fine threads.  On the other hand, panel (a), which is a cutout from the full-disk image, has the spatial resolution about 1--2$''$.  Therefore,  
the barbs seen here are understood to be groups of smaller threads below the spatial resolution.
The directions of the polarization signals in panel (b) coincide with the H$\alpha$ spine and barb structure very well. The direction somewhat varies from part to part of the filament, but it is approximately parallel to the H$\alpha$ spine or barbs in any part. This result confirms previous deductions that the structure in filaments is confined to follow the magnetic field. Plasma motions in filaments have been observed to be along the fine threads \citep{Lin2003, Lin2005, Lin2008}, and also verify the field-aligned nature of the threads of filaments.

Because the polarization directions of the filament in Figure \ref{fig:fig2}(b) are roughly aligned with either the spine or barbs, the average polarization direction for all of the filament structure without differentiating between the spine and barbs can be defined; it is shown by a dashed line in the inset at the lower-right of panel (b).
The green line shows the direction of the axis of the filament, which represents the average direction of the polarity inversion line seen in panel (b), which is approximately along a straight line.
In the case of the filament in Figure \ref{fig:fig2}, these two lines intersect at an angle of $18.0^\circ$.  In this case the average polarization direction deviates from the filament axis by $-18.0^\circ$.   The deviation angle is defined as an acute angle and a positive value corresponds to the
counterclockwise deviation; in this case, the deviation is clockwise.
This deviation angle 
is considered to be the deviation angle of the magnetic field vector with respect to 
the polarity inversion line.

As mentioned in Section 1, it is known that H$\alpha$ structures in filaments shows a chiral nature \citep[see e.g.,][]{Zirker1997, Martin1998}.  As in the case of the filament shown in Figure \ref{fig:fig2}, where 
the filament has barbs along the sides of the filament that bend downward to the right and make acute angles to the right side of the filament axis, the chirality is ``right-bearing'' (dextral).  
The alternative chirality is ``left-bearing'' (sinistral).  The spine/barbs structure and the deviation by an acute angle of the average polarization direction with respect to the axis of the filament are schematically illustrated at the bottom of Figure \ref{fig:fig2}.
It is known that right(left)-bearing filaments are dominant in the northern (southern) hemisphere, and actually the filament is located in the northern hemisphere as shown in Figure \ref{fig:fig1}.  
The magnetic field direction shown in Figure \ref{fig:fig2}(b) has the $180^\circ$ ambiguity, and it 
cannot be solved only with our polarization measurements.  However, 
 \citet{Martin1994} revealed that the direction of the magnetic field can be interpreted as shown in the illustrations at the bottom of Figure \ref{fig:fig2} \citep[it was discussed in detail by][]{Martin2008}.  The magnetic field in filament channels, which must exist above the polarity inversion line to form filaments, can be presumed based on the structure in fibrils around the polarity inversion line.  The magnetic field on either side of the channel is nearly parallel to the polarity inversion line, and points the same direction.  The majority of the filament channels show the dextral (sinistral) chirality in the northern (southern) hemisphere.  This magnetic field determines the magnetic field in the spine and barbs, as shown by orange arrows in the illustrations in Figure \ref{fig:fig2}.  Their direction is from the negative side to positive side, as previously found from the magnetic field measurement of prominences \citep[e.g.,][see review by \citeauthor{Bommier1998}\citeyear{Bommier1998}]{Leroy1983}.

As indicated above, analyzing measures of the average polarization orientation in filaments with respect to their axes is a new and practical means of learning more about the character of magnetic fields in filaments; taking into account that averaging could mask important small variations in field direction, averaging also can help to emphasize overall aspects of magnetic field configurations in a new statistical investigation.  In the following subsections, the method of the statistical study of the polarization orientation of filaments is described.

\subsection{Selection of observation days}

We picked filaments for the statistical study using the polarization maps taken on such selected days as
follows. First, we selected 89 days during 2010 April -- 2016 April
(including a nine month gap due to an instrumental failure caused by a lightning
strike) exactly every three weeks.  Filaments could survive during a couple of solar rotations, but the interval of three weeks enables us to avoid picking the identical
filaments repeatedly four or eight weeks (one or two solar rotations) apart.
If there
were no data on the selected day, we searched for the observation on a
different day around the selected day. From a set of the polarization maps taken on each
selected day, we picked filaments as below.

\subsection{Sample selection and derivation of polarization parameters}

\begin{figure}
\plotone{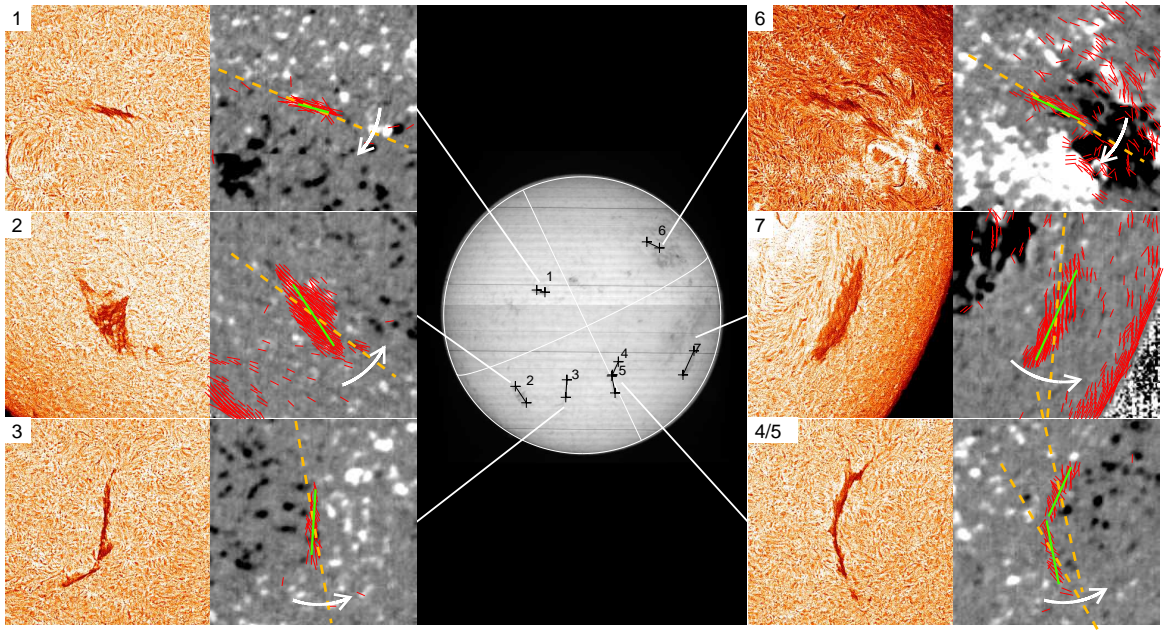}
\caption{
Filaments picked for the analysis and their polarization signals observed on 2015 October 27.  The \ion{He}{1}
1083.0 nm image at the center shows the positions of the picked filaments, and the enlarged views of the filaments are shown around it.  Each enlarged view ($6'.4\times6'.4$) shows the H$\alpha$ image of one of the filaments and its polarization signals.  The linear polarization signal of every $2\times2$ pixels (the degree of polarization
$>$ 0.1 \%) is shown with a red line. The average direction of the
linear polarization signals for each filament (or part of the filament) is shown with a dashed line. The background
gray scale image is a Stokes $V/I$ map of the \ion{Si}{1} 1082.7 nm line. The
approximate axis of each filament or part of filament is represented by a straight green line. 
\label{fig:fig3}}
\end{figure}

Figure \ref{fig:fig3}, which shows the observation on 2015 October 27, explains how the samples for the statistical study were selected and how their polarization parameters were derived for each observing day.  The derived polarization parameters and additional information for the samples in Figure \ref{fig:fig3} as well as the filament in Figures \ref{fig:fig1}/\ref{fig:fig2} are listed in Table \ref{tab:table1}.

(1) First we picked filaments remarkably visible in the \ion{He}{1} 1083.0 images and
define the axis of each filament.  Each axis is defined with a straight line, because it is the representative direction along which the filament extend, and it is compared with the average direction of polarization signals.  Since filaments often
meander on the solar surface, a filament is split into approximately linear
parts in such a case, and the axis is defined for each part. Picking
filaments is done manually by eye.
We did not pick filaments too close to the limb (farther than
about 55$^\circ$ from the disk center), to avoid the projection effect.
Filaments are picked regardless of whether they are active region filaments, quiescent filaments, or intermediate filaments (the class of the intermediate filaments, which form between weak unipolar background field regions and active region complexes, are often included in the classification of filaments \citep{Engvold1998}).
However, background plage regions of active region filaments
often show some polarization, and in such cases we do not pick them.
Therefore, the sampling of filaments is biased in favor of quiet and intermediate filaments, to some extent.

In the case of the 2015 October 27 data, six filaments are
picked; their positions are shown in the full-disk image of the \ion{He}{1} 1083.0 line in Figure \ref{fig:fig3}, and their enlarged views ($6'.4\times6'.4$) of the H$\alpha$ images and the polarization maps are shown around the full-disk image.  The axes of seven parts of the six filaments were defined as shown with the green lines in the polarization maps in Figure \ref{fig:fig3}.  
The spatial resolution of the H$\alpha$ images in Figure \ref{fig:fig3} is at most 1--2$''$. However, some of them show barbs and structures suggestive of unresolved fine threads.
Some details of the picked filaments, especially for the structures seen in the H$\alpha$ images, are described below. \\
Filament 1: This is a small filament in the northern hemisphere, showing the features of intermediate filaments, located not far from the disk center.  In the H$\alpha$ picture, dextral barb structures are recognized. \\
Filament 2: This filament had been high-lying from its appearance on the east limb, as often seen in a filament about to erupt; actually on the next day of the observation in Figure 3 it erupted.  Therefore, the filament is not completely quiescent, though it is located in the quiet area.  Such filaments are also automatically included for the statistical study. \\
Filament 3: This filament is another intermediate filament.  It is a curved filament in the H$\alpha$ image, but its lower-left part was faint in \ion{He}{1} 1083.0, probably due to the difference in the surrounding corona.  Therefore, we only pick the upper-right part for the study.  This filament is located in the southern hemisphere, and the H$\alpha$ structure suggests that this is sinistral. \\
Filament 4/5:  This is a curved filament in the southern hemisphere, and it is also an intermediate filament.  Because it is difficult to define the axis with a straight line, we divided it into two parts.  The barb structure seen in H$\alpha$ seems to be sinistral. \\
Filament 6: This filament is located close to an active region in the northern hemisphere.  The barb structure seen in H$\alpha$ seems to be dextral. \\
Filament 7:  This filament is located in the quiet area in the southern hemisphere.  Although the barbs are densely packed and the contrast is not high, the sinistral structure is recognized.

From the selected 89-day observations, we picked 376 filaments.  Some of 
them were divided into two or more parts; therefore the number of samples used 
for the study was 438.

(2) Next, we derived orientation of the linear polarization with respect to its axis for each
filament. If a filament was divided into some parts, the orientation of each
part was derived separately. 
In Figure \ref{fig:fig3}, polarization signals exceeding 0.1 \% are
shown with red lines in the enlarged polarization maps.
The polarization signals of every $2\times2$ pixels ($3''.5\times3''.5$) are shown as in Figure \ref{fig:fig1}.
These polarization signals are roughly aligned in each filament, and we can define the meaningful average polarization directions as for the filament in Figures \ref{fig:fig1}/\ref{fig:fig2}.
The average polarization directions are shown with dashed lines.  

On the basis of thus derived average polarizations and the axes of the filaments, the deviation angle of the average polarization from the axis is derived for each filament.  Filaments 1 and 6 show the clockwise deviation, and the other show the counterclockwise deviation with respect to their axes.
As described below, some of the filaments show the coincidence between the average direction of the polarization signals and the H$\alpha$ structure, as the filament in Figures \ref{fig:fig1}/\ref{fig:fig2} does. \\
Filament 1: The average magnetic field direction deviates clockwise from the axis, and it is consistent with the dextral barbs seen in H$\alpha$, as in the filament in Figures \ref{fig:fig1}/\ref{fig:fig2}. \\
Filament 2: Although the directivity of the H$\alpha$ structure is not clear, the average magnetic field direction of this filament, which is located in the southern hemisphere, was revealed to deviate counterclockwise. \\
Filament 3: The average magnetic field direction deviates counterclockwise, and it is consistent with the H$\alpha$ structure. \\
Filament 4/5:  Both the filament sections show that the average magnetic field direction deviates counterclockwise, and it is again consistent with the H$\alpha$ structure. \\
Filament 6: The magnetic field directions are nearly parallel to the filament axis, but their average direction deviates clockwise, even though the deviation angle is small.  The H$\alpha$ structure also seems to be dextral.\\
Filament 7:  The magnetic field directions are generally parallel to the barbs seen in H$\alpha$, and the average direction deviates counterclockwise. \\

\subsection{Derivation of auxiliary information}

In addition to the polarization parameters, some pieces of auxiliary
information were also derived as below.

(1) Heliographic positions of the center of the filaments were derived. Based on the derived positions, the apparent deviation angles
of the polarization derived above (measured on the sky plane)
were converted to the true deviation angles of the magnetic field vector from the (true) filament axis, on the assumption that both the
filament axis and the magnetic field vector are horizontal.  In Table \ref{tab:table1}, the heliographic (Stonyhurst) positions of the filaments in Figures \ref{fig:fig1}/\ref{fig:fig2} and \ref{fig:fig3} and the apparent and true deviation angles of their magnetic field vectors are listed.  For instance, the true deviation angle of the filament in Figures \ref{fig:fig1}/\ref{fig:fig2}, of which the apparent deviation angle of was derived to be $-18.0^\circ$, is calculated to be $-19.6^\circ$.  The filaments being not close to the disk center show considerable differences between the apparent and true deviation angles.  The true deviation angles are used for the statistical study.

(2) We divided the location of the filaments into quiet and active areas.
For simplicity, we did not adopt the class of the intermediate filaments.
The active areas in our classification include not only inside active regions but also around their outskirts,
because network structures brighter than those in quiet regions in H$\alpha$
surround active regions.  Therefore, the intermediate filaments are partly classified into the filaments in the active areas, while most of them are defined to be in the quiet areas.
This classification was done by eye. The filament in Figures \ref{fig:fig1}/\ref{fig:fig2} is located in a quiet area.  In the case of filaments in Figure
\ref{fig:fig3}, filaments 1, 2, 3, 4/5, and 7 are defined to be located in quiet areas.
Filament 6 is located in the vicinity of an active region, and such a case is
classified as a filament located in an active area.

(3) We classified the alignment of the magnetic polarities in the background photosphere of
the filaments.  According to the Hale-Nicholson law, the magnetic polarities of active regions are known to have the same alignment in a hemisphere, and that in the northern hemisphere and that in the southern hemisphere are opposite to each other.  
Furthermore, the polarity alignment switches every solar cycle; in the current solar cycle 24, the negative (positive) polarity precedes, or is located westward, in the northern (southern)
hemisphere.  On the other hand, filaments appear above the magnetic polarity boundaries, regardless of whether the alignment is the same as or the opposite to ordinary bipolar regions.  If the preceding polarity in the background magnetic field of a filament is negative (positive) in the northern (southern) hemisphere, the alignment is defined to be the Hale type, because
the magnetic polarities is the same as that of ordinary bipolar regions in the corresponding hemisphere.  In the case of reversed polarities,
the alignment is defined to be the non-Hale type.  In fact, it is known that more filaments are formed at the non-Hale type border than at the Hale type border \citep{Tang1987}.
In the case of filaments, the preceding polarities are not necessarily located to the west, but  sometimes located to their poleward side, unlike ordinary bipolar regions \citep[see e.g., Figure 9 of][as a helpful illustration]{Rust1967}.

In Figure \ref{fig:fig2}(b) and the enlarged polarization maps in Figure 
\ref{fig:fig3}, the background images show the longitudinal photospheric magnetic field (white and black correspond to the positive and negative polarities).    In Figure \ref{fig:fig2}(b), the positive (negative) polarity area is located to the north (south) of the filament as shown in the inset of panel (b).  In this case, the positive polarity is preceding in the northern hemisphere, and therefore, this one is defined to be the non-Hale type.
In Figure \ref{fig:fig3}, positive polarity area precedes for filaments 1 and 3, and negative
polarity area precedes for filament 2, 4, 5, and 6. Then the general east-west alignment of the background photospheric magnetic polarities
of filaments 3 (southern hemisphere) and 6 (northern hemisphere) is the same as that of ordinary bipolar regions in the corresponding hemisphere, and such cases are defined to be the Hale type.  On the other hand, that of
filaments 1, 2, 4 and 5 is defined to be the non-Hale type. The polarity for
filament 7 is not clear. Such a case is classified as ``unknown''.

The derived parameters for the filaments shown in Figures \ref{fig:fig1}/\ref{fig:fig2} and \ref{fig:fig3} are also listed in Table \ref{tab:table1}.  For all the 438 samples, we derived such parameters for the statistical study.

\begin{table}
\caption{Polarization Properties and Auxiliary Information of Sample Filaments \label{tab:table1}}
\begin{tabular}{lrrrrrccc}
\hline\hline
 & & \multicolumn{2}{c}{Heliographic Position} & \multicolumn{2}{c}{Deviation Angle} & & & \\
Date & No. & Longitude & Latitude & Apparent & True & Quiet/Active & Preceding Polarity & Magnetic Alignments \\
\hline
2014 Nov 23 &  &    11.6  &    22.5  &   -18.0  &   -19.6  & Quiet  &    P & non-Hale \\
2015 Oct 27 & 1 &    -12.3  &     20.8  &   -7.3  &   -6.9  & Quiet  &   P & non-Hale \\
2015 Oct 27 & 2 &   -42.6  &     -16.3  &    19.4   &   27.5   & Quiet  &     N & non-Hale \\
2015 Oct 27 & 3 &    -21.1 &     -21.6 &     13.6  &    11.6 & Quiet  &    P & Hale \\
2015 Oct 27 & 4 &    3.1 &     -22.4  &     37.3  &    36.1   & Quiet  &    N & non-Hale \\
2015 Oct 27 & 5 &   -0.8  &     -29.1 &    19.2  &    16.0  & Quiet  &    N & non-Hale \\
2015 Oct 27 & 6 &   45.1  &      16.9   &   -5.7 &    -7.5   & Active  &    N & Hale \\
2015 Oct 27 & 7 &    43.1   &     -37.1  &   19.5  &    33.8  & Quiet  & ? & unknown \\
\hline\hline
\multicolumn{9}{l}{Notes. Various kinds of parameters of the filaments shown in Figures {\ref{fig:fig1}/\ref{fig:fig1} and \ref{fig:fig3}}.}\\
\end{tabular}
\end{table}

\section{Results} \label{sec:results}

\begin{figure}
\plotone{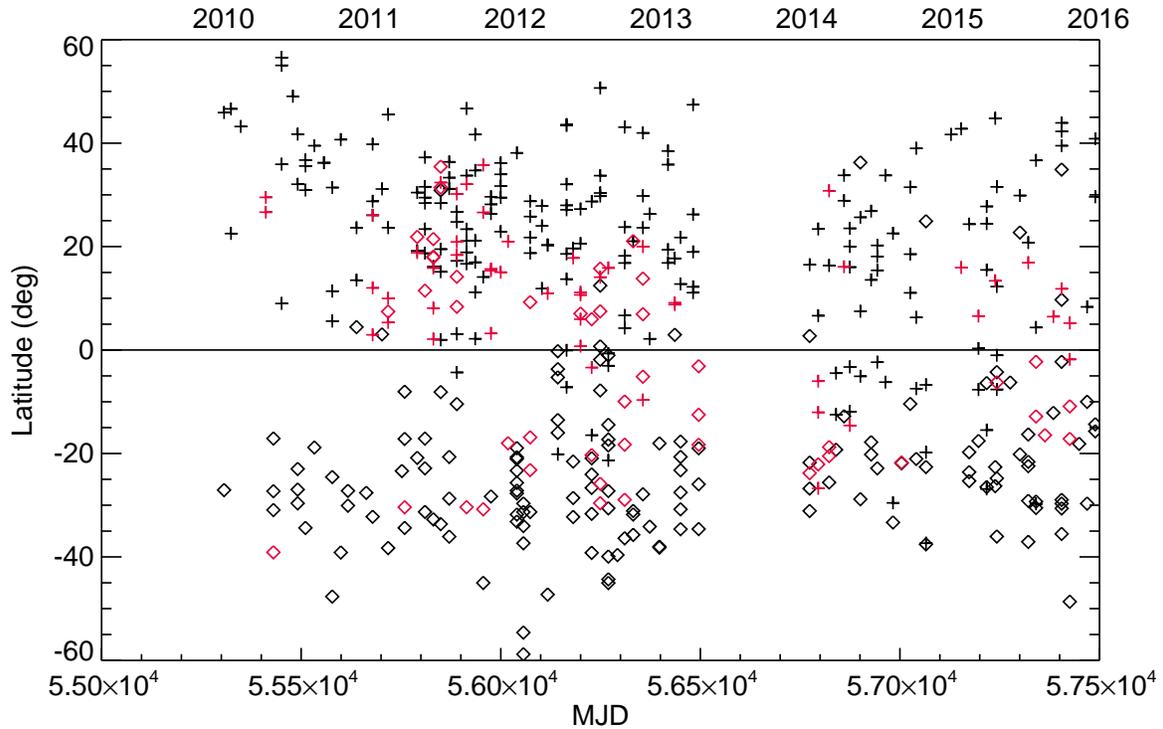}
\caption{Time-versus-latitude plot of the selected filaments. The time is
depicted both in the modified Julian day (bottom axis) and the year (top axis). Plus symbols show the
filaments whose average polarization orientation
deviates clockwise with respect to the filament axis, and
diamonds show that  
deviates counterclockwise. Black symbols represent filaments in quiet areas, and
red symbols represent those in active areas. \label{fig:fig4}}
\end{figure}

\begin{table}
\caption{Number of Filament Samples \label{tab:table2}}
\begin{tabular}{lrrr}
Quiet Area & & & \\
\hline\hline
 & consistent & inconsistent & total \\
\hline
Hale & 88 & 3 & 91 \\
Non-Hale & 151 & 27 & 178 \\
unknown & 65 & 8 & 73 \\
\hline
total & 304 & 38 & 342 \\
\hline\hline
&&&\\
Active Area & & & \\
\hline\hline
 & consistent & inconsistent & total \\
\hline
Hale & 30 & 8 & 38 \\
Non-Hale & 26 & 13 & 39 \\
unknown & 15 & 4 & 19 \\
\hline
total & 71 & 25 & 96 \\
\hline\hline
\multicolumn{4}{p{20em}}{Notes. The numbers of filament samples divided by their characteristics; located in quiet 
or active areas, consistent or inconsistent with the hemispheric pattern, and 
having Hale, non-Hale, or unknown magnetic configurations.}\\
\end{tabular}
\end{table}

\begin{figure}
\plotone{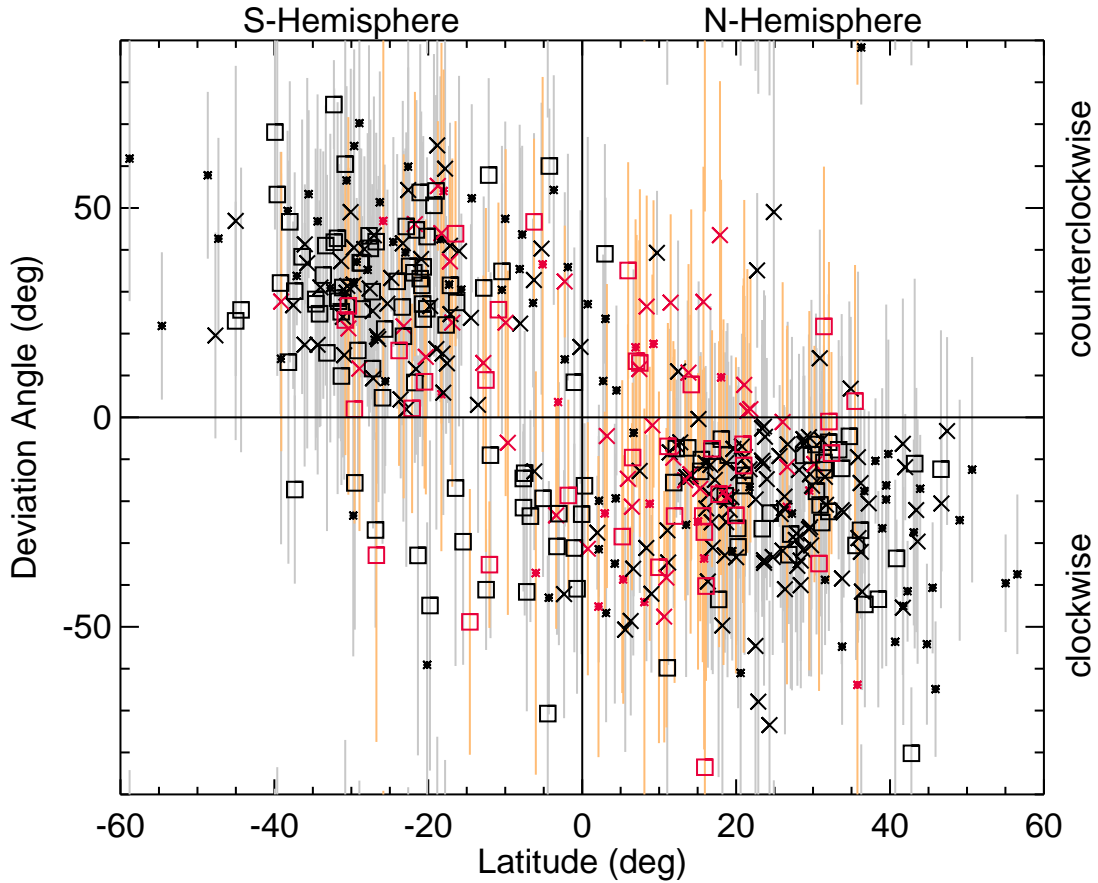}
\caption{The relation between the latitude of the filaments and their
average deviation angle. Black symbols represent filaments in the quiet
areas, and red symbols represent those in the active areas. The cross symbols correspond to the
filaments with the preceding positive polarity, and the box symbols corresponds to the filaments with the preceding
negative polarity. The dots show the filaments where the alignment of the polarity
cannot be defined clearly.  The vertical lines are error bars, showing the $\pm1\sigma$ range of the distribution of the deviation
angles before the averaging in each filament.  The gray and orange bars correspond to the filaments in the quiet and active areas. \label{fig:fig5}}
\end{figure}

\begin{figure}
\plotone{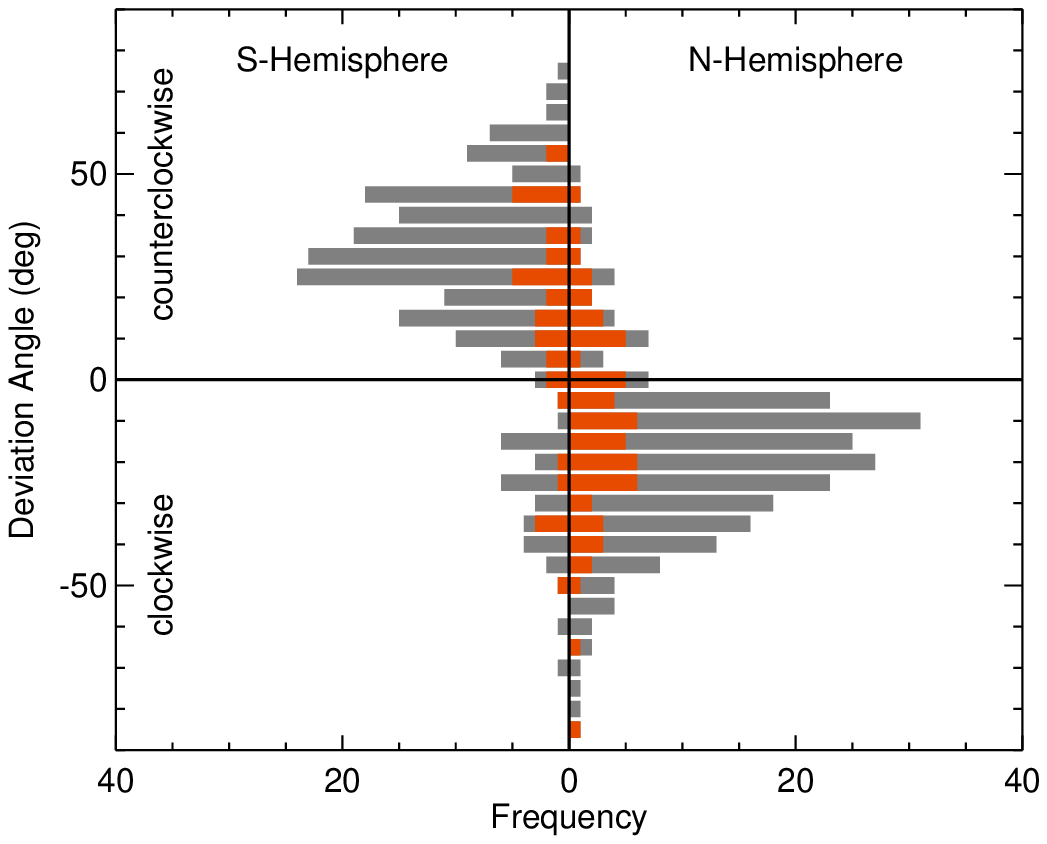}
\caption{Histogram of the average deviation angle of filaments
with respect to the filament axis presented separately for the two hemispheres.  Filaments in quiet and active areas shown with gray and orange bars are stacked. \label{fig:fig6}}
\end{figure}

The results of the statistical analysis is summarized in Figures \ref{fig:fig4}, \ref{fig:fig5}, and \ref{fig:fig6}, and Table \ref{tab:table2}.  Their descriptions are given below.

\subsection{Average orientation of the filament magnetic field with respect to the filament axis}

Table \ref{tab:table1} shows that the filaments in the northern hemisphere, including the one on 2014 November 23 shown in Figures \ref{fig:fig1} / \ref{fig:fig2} and filaments 1 and 6 on 2015 October 27 in Figure \ref{fig:fig3}, show
the clockwise deviation of the average polarization orientation with respect to the
filament axis (negative deviation angle), and those in the southern hemisphere show
the counterclockwise deviation (positive deviation angle). These results suggest that there is a systematic property in the deviation angle, depending on the hemisphere where the filaments appear.  This hemispheric pattern of the sense of the
deviation (i.e. chirality) was investigated for all the selected filaments from the statistical viewpoint, 
and the results are as follows.

Figure \ref{fig:fig4} is the time-versus-latitude plot of all the selected filaments. This is a
kind of the butterfly diagram of filaments, but note that it lacks filaments close
to the poles, because the filaments close to the limb were excluded in this statistical study.  Black and red symbols
correspond to the filaments in the quiet areas and those in the active areas, respectively.
The
plus symbols show the filaments whose average direction of polarization deviated clockwise
with respect to the filament axis; the plus symbols are concentrated in the northern hemisphere.
On the other hand, the diamonds, corresponding to the filaments with counterclockwise 
deviation, are concentrated in the southern hemisphere.  
In summary, there is a strong correlation between hemisphere and clockwise or counterclockwise sense of deviation in the acute angles of the average direction of the magnetic field in filaments from the axis of the filaments. This hemispheric pattern is consistent with previous findings for the statistical associations between hemisphere and the chiralities of solar features.

In Table \ref{tab:table2}, the numbers of filaments that are consistent or inconsistent with the
hemispheric pattern are listed, dividing them in terms of their location (quiet or active) and
magnetic polarity alignment of the photospheric magnetic fields next to filaments (Hale, non-Hale, and unknown), to investigate the
hemispheric tendency more quantitatively. For the definitions of the location and the magnetic polarity alignment, see Section 2.4.
In the quiet areas, the number of the non-Hale type filaments are almost twice of that of the Hale type ones, as found by \citet{Tang1987}.
The samples of the dominant sense of the deviation of the polarization direction from the filament axis (clockwise in the northern
hemisphere/ counterclockwise in the southern hemisphere) occupy 89 \% (304
samples) in the 342 quiet area samples and 74 \% (71 samples) in the 96
active area samples. There is a tendency for the
filaments in the quiet area to more strongly show the hemispheric pattern (but note that quiescent filaments near the poles and many active region filaments are not included in our study).
This hemispheric pattern of the chirality is kept throughout the six-year period. This period covers most of the active phase of solar cycle 24.
Taking a closer look, we can find that the quiet area samples which are inconsistent with the hemispheric pattern increased from 6 \% (14 out of 227 samples) before the observation gap in 2013 to 21 \% (24 out of 115) after the gap.

Figure \ref{fig:fig5} shows the relation between the latitude of the filaments and their
average deviation angle. Black and red symbols again
correspond to the filaments in the quiet areas and those in the active areas, respectively. 
The average deviation angle of an individual filament is determined on 
the basis of the average linear polarization signal, but the polarization signals 
in a filament have some dispersion.  The error bars (the vertical lines) show the $\pm1\sigma$ 
range of the distribution of the deviation
angle in each filament. Again, we can find that the clockwise deviation (negative deviation angle) is dominant in the northern
hemisphere, and the counterclockwise deviation (positive deviation angle) is
dominant in the
southern hemisphere. 

Furthermore, the average deviation angles are concentrated in a limited range; in
both hemispheres, there are only a small number of filaments with large absolute
deviation angles ($>50^\circ$), or deviation angles
close to $0^\circ$. The deviation angle seems to be concentrated in a
certain range regardless of the latitude. Figure \ref{fig:fig6} shows a histogram of the
average deviation angle.   The distribution of the deviation angle has its peak around
10--30$^\circ$ in both hemispheres. Note that the deviation angle of 90$^\circ$
means that the magnetic field is perpendicular to the filament axis. 
In Figure \ref{fig:fig6}, the filaments in the
quiet areas and those in the active areas are shown with gray and red bars. Figure \ref{fig:fig6} shows that if we see the filaments in the active areas only,
there are a small number of filaments which have large deviation angles, but the
concentration of the deviation angles within the limited range is not very clear.
The concentration in the 10--30$^\circ$ range is one of the characteristics particularly of
the filaments in the quiet areas.

\subsection{Alignment of the magnetic polarities in the background photosphere of the filaments}

As shown in Figures \ref{fig:fig2} and \ref{fig:fig3}, in the background network magnetic fields on both sides of the
filaments, the positive polarity precedes in some cases, and the negative
polarity precedes in the other cases. As explained in Section 2.4, the preceding polarity is placed on the western or the poleward side of the inversion line.  Figure \ref{fig:fig5} includes information on the
background magnetic field of each sample. The cross symbols correspond to the
filaments with preceding positive polarity, and the box symbols correspond to the filaments
with preceding negative polarity. The dots show the filaments where the polarity
alignment cannot be defined clearly. 
As described in Section 2.4, according to the Hale-Nicholson law, the negative (positive) polarity precedes in the northern (southern) hemisphere in active regions in solar cycle 24.  Therefore, in the northern hemisphere, the box symbols are the Hale type samples and the cross symbols are the non-Hale type ones, and the relation is opposite in the southern hemisphere.
As seen in the upper-left and the lower-right sections
of Figure \ref{fig:fig5} (filaments being consistent with the hemispheric pattern), both of the magnetic polarity alignments are mixed,
regardless of whether they were in the quiet area samples or the active area
samples. 
On the other hand, in filaments that are not consistent with the hemispheric pattern (the upper-right and
the lower-left sections in Figure \ref{fig:fig5}), the non-Hale type samples (crosses in
the upper-right section and boxes in the lower-left sections) are noticeable. Actually, as
shown in Table \ref{tab:table2}, filaments in the quiet area particularly show a remarkable
imbalance; among the filaments not being consistent with the hemispheric pattern, there are 27 non-Hale samples and only three Hale samples. In the entire quiet area samples (including both the consistent and inconsistent ones with the hemispheric pattern), the number of
non-Hale samples, 178, is nearly twice of the number of Hale samples, 91 as mentioned in Section 3.1. Even
so, it is worth noting that the filaments which are not consistent with the hemispheric pattern rarely appear under
the Hale type magnetic polarity alignment. The filaments in the active area show the
same tendency; eight out of 38 Hale samples and 13 out of 39 non-Hale sample
are not consistent with the hemispheric pattern. In these numbers, some filaments are
counted multiple times, because a part of the filaments were divided into
portions. However, even if we take this into account, the conclusion does not
change at all.

\section{Discussion} \label{sec:discussion}

\subsection{Average magnetic field direction in filaments}

As mentioned in Section 1, fine structures in filaments are well known to
show the hemispheric pattern \citep[e.g.,][]{Zirker1997, Martin1998}; furthermore, the observations of 
magnetic field in prominences previously 
showed the same type of the hemispheric pattern \citep[reviewed by][]{Bommier1998}.  
Our result that the average magnetic field
orientation of individual filaments deviates clockwise/ counterclockwise in the northern/ southern
hemisphere with respect to the axes of filaments is consistent with these former results. 
Therefore, our result confirmed the general characteristics of the magnetic field
orientation in filaments with the direct observations of the
magnetic field in filaments for the first time.
Although the H$\alpha$ images in Figures {\ref{fig:fig1}/\ref{fig:fig2} and \ref{fig:fig3}} are not very high resolution ones, we can confirm that in some of the filaments the direction of the polarization signals is actually parallel to the structures seen in the H$\alpha$ filaments.  From comparison with formerly published high resolution observations, it is deduced that the magnetic field is aligned with the unresolved fine threads seen in H$\alpha$ images.  A detailed comparison with higher spatial resolution H$\alpha$ data taken simultaneously with the magnetic field data will be a future work.

Our results that 89 \% of the filaments in the quiet areas and 74 \% of those in the active areas follow the hemispheric pattern are quantitatively consistent with those from previous studies of the structure in filaments observed in H$\alpha$ images \citep[e.g.,][]{Martin1994, Pevtsov2003} and in extreme ultraviolet images \citep{Ouyang2017}.  \citet{Ouyang2017} pointed out
the increase in the quiet area filaments which are inconsistent with the hemispheric pattern during the period of 2010--2015, and this tendency can also be found in our results as mentioned in Section 3.1.  This coincidence also substantiates the similarity between the fine structures and the magnetic field vectors in filaments.
Note that a certain number of filaments which do not show the hemispheric pattern have been found in all the studies.
The models of the formation of the hemispheric pattern proposed by \citet{Mackay2005} and \citet{Yeates2009} succeeded to reproduce the exceptions as well.  

Our sample selection is biased to some extent against the filaments in the active areas.  It is difficult to isolate the polarization signals in small, narrow filaments in active regions and those located in plage regions from the background.  Therefore, the results for the filaments in the active areas should be recognized to be qualitative to some extent.  On the other hand, the quiet area filaments have been selected without bias except that the polar crown filaments have not been picked due to their location being too close to the limb.  This fact means that our results for the filaments in the quiet areas represent the characteristics of the ordinary quiescent (and also intermediate) filaments.  Another possible bias is that filaments which cannot be remarkably seen in the \ion{He}{1}
1083.0 line have not been selected.  Visibility of filaments sometimes depends on the wavelength \citep[e.g., see][for the comparison between images in the \ion{He}{2} 30.4 nm line and H$\alpha$]{Martin2015}, and it is necessary that the filaments show enough absorption both in the \ion{He}{1}
1083.0 and H$\alpha$ lines for our study.

Our observations show that the deviation angles of the observed magnetic field
in filaments from their axis, particularly those in the quiet areas, are concentrated in a limited
angle range with a peak around 10--30$^\circ$.
From the magnetic field observations of prominences,  \citet{Tandberg-Hanssen1970} deduced that the orientation of the magnetic field makes an angle of about
15$^\circ$ with respect to the long axis of the prominences, and 
\citet{Bommier1998}
showed that this angle is about
40$^\circ$. These results can be interpreted as that a certain deviation of the magnetic field direction with respect to
the polarity inversion line is necessary to support a filament stably.

As a whole, it is a possible conjecture based on the observational results of the magnetic field of filaments 
that the filaments, particularly those in the quiet areas, are formed above
the magnetic polarity inversion line where shear of a preferable sign is
produced by the evolution of the global magnetic field of the Sun. 
In particular, at the polarity inversion line with the Hale type magnetic
polarity alignment, it is likely that the magnetic shear which follows the hemispheric pattern is
efficiently produced. 
On the other hand, the reason why the filaments in the
active area do not show the hemispheric pattern very much can be presumed that
the background magnetic field of filaments in the active
area is more influenced by the evolution of the magnetic field of active
regions than the global magnetic evolutions.

\subsection{Implication to the CME studies}

\begin{figure}
\epsscale{0.5}
\plotone{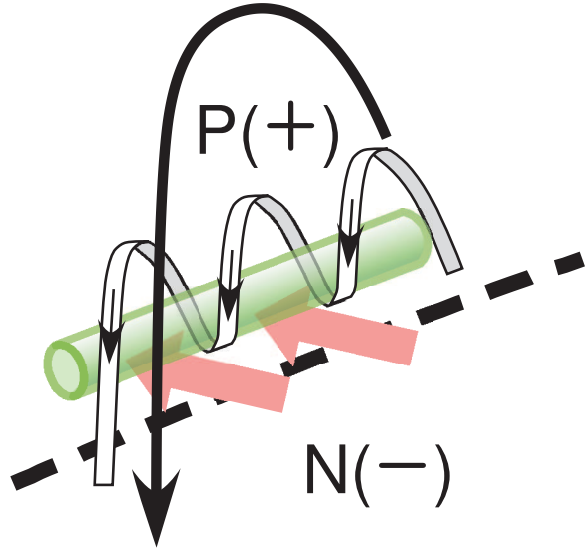}
\caption{Schematic drawing of a filament and surrounding magnetic field. There
is a flux rope above a magnetic polarity inversion line (dashed line), and
filament material (green) at the bottom of the flux rope. The observed
magnetic field in filaments corresponds to those shown with red arrows. The magnetic field in the corona
shown by an arrow above the flux rope is also strongly sheared, and its sense
of deviation from the polarity inversion line is opposite to that of the filament. \label{fig:fig7}}
\end{figure}

Our study is limited to the filaments, but we would like to call the readers' attention to its relation to the magnetic field in the corona surrounding the filaments and its implication to the CMEs.
Besides the hemispheric pattern of the magntic field in filaments, 
it is known that the handedness of the skew of coronal loops is apparently opposite to that
of filaments as mentioned in Section 1. 

As also mentioned in Sections 1 and 2.1, the magnetic field in filaments show the ``inverse'' characteristics that
the magnetic field is directed from the negative polarity
side to the positive \citep{Bommier1998, Martin1994}.
\citet{Chen2014} proposed to use extreme ultraviolet images to resolve the $180^\circ$ ambiguity; on the basis of the skew of the footpoints of the magnetic field of the filament, they concluded that the magnetic field of the filament was found to be consistent with the above results.  
\citet{Martin1998} attributed the cause of the inverse magnetic field to small opposite polarity patches in the background magnetic field of filaments \citep[the three-dimensional structure of the complete chiral system including the opposite polarity patches is illustrated in][]{Martin2012}.
On the other hand, the inverse magnetic field is known to be understandable on the assumption that the filament material is located above the X-type neutral point of the magnetic field, as originally suggested by \citet{Kuperus1974}.  The formation of such a magnetic field structure was discussed by \citet{vanBallegooijen1989, vanBallegooijen1990} and \citet{RustKumar1994}, and they suggested that a filament is located at the bottom of a
flux rope, which has a helical magnetic field \citep[see also][]{vanBallegooijen2014}.  
In any case, the opposite handedness between the filament and the coronal magnetic field is consistently understandable.

The inverse magnetic field and the opposite handedness give important information of the magnetic field in CMEs.  A flux rope formed above the polarity inversion line occasionally erupts and becomes a CME
\citep[e.g.,][]{Marubashi1997}.  The structure of a flux rope in a CME is
observed in the interplanetary space with in situ magnetic field measurements, and
the relation between the handedness of the magnetic field in CME flux ropes
and the hemisphere of their origin was known as mentioned in Section 1.
According to the model of the flux rope with a helical magnetic field mentioned above, the magnetic field in filaments and surrounding corona can be represented by a cartoon in Figure
\ref{fig:fig7}.
The magnetic field in the corona, which is somewhat sheared, and
that of the upper half of the flux rope are parallel, but that of the lower
half of the flux rope, where filaments are located, has a different orientation. 
Furthermore, the magnetic field
in the lower half of the flux rope seems to be directed from the negative
polarity to the positive polarity.  
It is controversial if the flux rope structure is inherent in filaments in any stage or formed as the result of magnetic reconnection leading to the eruption.  
The structure shown in Figure
\ref{fig:fig7} can be considered to be formed before CMEs.  
Alternatively, in the \citet{Martin2012} concept, the flux rope that becomes the CME, forms suddenly during the same magnetic reconnection that produces the associated flare. In either case, the flux rope that forms
eventually erupts into the interplanetary space, and the magnetic field at the bottom of the flux rope (red arrows in Figure \ref{fig:fig7}) is considered to evolve into the magnetic field of the rear side  of the interplanetary flux rope.  It means that the magnetic field in interplanetary clouds, which sometimes causes geomagnetic storms, can be known before the eruption with the observation of the filament magnetic field.  

This simple view does not explain some aspects of the filaments such as barbs connected to the chromosphere and material supply into the filaments.
However, the model in Figure 7 is supported by recent observations.
Some observations by the Atmospheric Imaging Assembly of the Solar Dynamics
Observatory, which show the details of the structures of the
lower corona, show that filaments are located at the bottom of coronal cavity
\citep[e.g.,][]{Berger2012, Regnier2011}. \citet{Su2012} and
\citet{MartinezGonzalez2015}.  These authors studied the magnetic field in filaments,
showed that helical magnetic fields surround the filaments.

There is another topic to be mentioned here.  It has been pointed out that filaments with the Hale-type magnetic
field alignment tend to be related to more CMEs than those
with the non-Hale type alignment \citep{Marubashi2015}. Therefore, the result
that the filaments with the Hale-type magnetic alignment more strongly show
the hemispheric pattern will be a clue to investigate what kind of structure and
evolution of the magnetic field in filaments are related to CMEs.

\subsection*{}

Such a statistical study will contribute to understanding of the conditions of
the magnetic field where filaments are formed. In addition, the magnetic field
of the individual filament and its evolution can be investigated with our
observation. Polarizations in prominences can be also measured, and they give information of the magnetic field as well.  Therefore, our data, combined with the high-precision
photospheric magnetic field data and information of the coronal loop
structures, will contribute to understanding of the process of filament
formation and eruption. Furthermore, if high spatial resolution data taken at various positions on the solar disk are available,  detailed studies of inidividual filaments will become possible, which clarify the relation between the fine structures in H$\alpha$ and their magnetic field.
Such studies will eventually reveal the three-dimensional structure of the magnetic field in and around filaments.  In the case that the filaments are in the activation phase, the magnetic field of CMEs, which are extending to the interplanetary space, can be presumed on the basis of the observed magnetic field in filaments,
even if the coronal loop structures observed in the X and EUV wavelengths is not very clear.

\acknowledgments

This work is partly supported by JSPS KAKENHI Grant Number JP17204014 and JP15H05814, and also by NAOJ research grant.  The authors are grateful to the anonymous referee for careful reading and excellent, detailed comments.


\begin{thebibliography}{}

\bibitem[Berger(2012)]{Berger2012}
Berger, T. 2012, ASPC 463, Magnetic Fields from the Photosphere to the Corona, ed T. Rimmele, A. Tritschler, F. W\"oger et al. (San Francisco: ASP), 147

\bibitem[Bernasconi et al.(2005)]{Bernasconi2005} Bernasconi, P.~N., Rust, D.~M., \& Hakim, D.\ 2005, \solphys, 228, 97 

\bibitem[Bommier et al.(1994)]{Bommier1994}
Bommier, V., Landi Degl'Innocenti, E., Leroy, J.-L., \& Sahal-Brechot, S.\ 1994, \solphys, 154, 231 

\bibitem[Bommier \& Leroy(1998)]{Bommier1998}
Bommier, V. \& Leroy, J.L. 1998, ASPC 150, New Perspective on Solar Prominences, ed D.F. Webb, B. Schmieder, \& D.M. Rust (San Francisco: ASP), 434

\bibitem[Bommier et al.(1986)]{Bommier1986}
Bommier, V., Sahal-Brechot, S., \& Leroy, J.~L.\ 1986, \aap, 156, 79 

\bibitem[Bothmer \& Schwenn(1994)]{Bothmer1994}
Bothmer, V., \& Schwenn, R.\ 1994, \ssr, 70, 215 

\bibitem[Chen et al.(2014)]{Chen2014} Chen, P.~F., Harra, L.~K., \& Fang, C.\ 2014, \apj, 784, 50 

\bibitem[Engvold(1998)]{Engvold1998} Engvold, O.\ 1998, IAU Colloq.~167: New Perspectives on Solar Prominences, 150, 23 

\bibitem[Foukal(1971)]{Foukal1971} Foukal, P.\ 1971, \solphys, 19, 59 

\bibitem[Hagino \& Sakurai(2005)]{Hagino2005} Hagino, M., \& Sakurai, T.\ 2005, \pasj, 57, 481 

\bibitem[Hale(1927)]{Hale1927} Hale, G.~E.\ 1927, \nat, 119, 708 

\bibitem[Hanaoka et al.(2011)]{Hanaoka2011}
Hanaoka, Y., Sakurai, T., Shinoda, K., et al.  2011, ASPC 150, Solar Polarization 6, ed J. R. Kuhn, D. M. Harrington, H. Lin et al. (San Francisco: ASP), 371

\bibitem[Hanaoka \& Sakurai(2014)]{Hanaoka2014}
Hanaoka, Y., \& Sakurai, T.  2014,
IAUS 300, Nature of Prominences and their role in Space Weather, ed. B. Schmieder, J.-M. Malherbe, \& S.T. Wu, 515 

\bibitem[Kuckein et al.(2009)]{Kuckein2009}
Kuckein, C., Centeno, R., Mart\'inez Pillet, V., Casini, R., Manso Sainz, R., \& Shimizu, T.
2009, \aap, 501, 1113

\bibitem[Kuperus \& Raadu(1974)]{Kuperus1974} Kuperus, M., \& Raadu, M.~A.\ 1974, \aap, 31, 189 

\bibitem[Leroy et al.(1983)]{Leroy1983}
Leroy, J.~L., Bommier, V., \& Sahal-Brechot, S.\ 1983, \solphys, 83, 135 

\bibitem[Leroy et al.(1984)]{Leroy1984}
Leroy, J.~L., Bommier, V., \& Sahal-Brechot, S.\ 1984, \aap, 131, 33 

\bibitem[Lim \& Chae(2009)]{Lim2009} Lim, E.-K., \& Chae, J.\ 2009, \apj, 692, 104 

\bibitem[Lin et al.(2003)]{Lin2003} Lin, Y., Engvold, O., \& Wiik, J.~E.\ 2003, \solphys, 216, 109 

\bibitem[Lin et al.(2005)]{Lin2005} Lin, Y., Engvold, O., Rouppe van der Voort, L., Wiik, J.~E., \& Berger, T.~E.\ 2005, \solphys, 226, 239 

\bibitem[Lin et al.(2008)]{Lin2008} Lin, Y., Martin, S.~F., \& Engvold, O.\ 2008, ASPC 383, Subsurface and Atmospheric Influences on Solar Activity, ed R. Howe, R.~W. Komm, K.~S. Balasubramaniam \& G.~J.~D. Petrie (San Francisco: ASP), 235 

\bibitem[Mackay \& van Ballegooijen(2005)]{Mackay2005} Mackay, D.~H., \& van Ballegooijen, A.~A.\ 2005, \apjl, 621, L77 

\bibitem[Martin et al.(1994)]{Martin1994}
Martin, S. F., Bilimoria, R., \& Tracadas, P. W. 1994, in NATO Advanced
Science Institutes (ASI) Series C, Vol. 433, ed. R. J. Rutten \&
C. J. Schrijver (Dordrecht: Kluwer), 303

\bibitem[Martin(1998)]{Martin1998}
Martin, S. F. 1998, ASPC 150, New Perspective on Solar Prominences, ed D.F. Webb, B. Schmieder, \& D.M. Rust (San Francisco: ASP), 419

\bibitem[Martin(2015)]{Martin2015} Martin, S.~F.\ 2015, Solar Prominences, 415, 205 

\bibitem[Martin et al.(2008)]{Martin2008} Martin, S.~F., Lin, Y., \& Engvold, O.\ 2008, \solphys, 250, 31 

\bibitem[Martin \& McAllister(1996)]{Martin1996} Martin, S.~F., \& McAllister, A.~H.\ 1996, IAU Colloq.~153: Magnetodynamic Phenomena in the Solar Atmosphere - Prototypes of Stellar Magnetic Activity, 497 

\bibitem[Martin et al.(2012)]{Martin2012} Martin, S.~F., Panasenco, O., Berger, M.~A., et al.\ 2012, Second ATST-EAST Meeting: Magnetic Fields from the Photosphere to the Corona., 463, 157 

\bibitem[Mart\'inez Gonz\'alez et al.(2015)]{MartinezGonzalez2015}
Mart\'inez Gonz\'alez, M. J., Manso Sainz, R., Asensio Ramos, A., et al. 2015, \apj, 802, 3

\bibitem[Marubashi(1997)]{Marubashi1997}
Marubashi, K. 1997, Coronal Mass Ejections, Geophys. Monogr. Ser., 99, ed N. Crooker, J. A. Joselyn, \& J. Feynman,  (Washington, D. C.: AGU), 147

\bibitem[Marubashi et al.(2015)]{Marubashi2015}
Marubashi, K., Akiyama, S., Yashiro, S., et al.\ 2015, \solphys, 290, 1371 

\bibitem[Ouyang et al.(2017)]{Ouyang2017}
Ouyang, Y., Zhou, Y. H., Chen, P. F., \& Fang C. 2017, \apj, 835, 94

\bibitem[Orozco Su\'arez et al.(2014)]{OrozcoSuarez2014}
Orozco Su\'arez, D., Asensio Ramos, A., \& Trujillo Bueno, J.
2014, \aap, 566, A46

\bibitem[Pevtsov et al.(2003)]{Pevtsov2003} Pevtsov, A.~A., Balasubramaniam, K.~S., \& Rogers, J.~W.\ 2003, \apj, 595, 500 

\bibitem[Pevtsov et al.(1995)]{Pevtsov1995} Pevtsov, A.~A., Canfield, R.~C., \& Metcalf, T.~R.\ 1995, \apjl, 440, L109 

\bibitem[R\'egnier et al.(2011)]{Regnier2011}
R\'egnier, S., Walsh, R. W., \& Alexander, C. E 2011, \aap, 533, L1

\bibitem[Richardson(1941)]{Richardson1941} Richardson, R.~S.\ 1941, \apj, 93, 24 

\bibitem[Rust(1967)]{Rust1967} Rust, D.~M.\ 1967, \apj, 150, 313 

\bibitem[Rust(1994)]{Rust1994}
Rust, D.~M. 1994, \grl, 21, 241

\bibitem[Rust \& Kumar(1994)]{RustKumar1994} 
Rust, D. M. \& Kumar, A. 1994, \solphys, 155, 69

\bibitem[Rust \& Kumar(1996)]{Rust1996} Rust, D.~M., \& Kumar, A.\ 1996, \apjl, 464, L199 

\bibitem[Sakurai et al.(1995)]{Sakurai1995}
Sakurai, T., Ichimoto, K., Nishino, Y., et al. 1995,
\pasj, 47, 81

\bibitem[Sheeley et al.(2013)]{Sheeley2013} Sheeley, N.~R., Jr., Martin, S.~F., Panasenco, O., \& Warren, H.~P.\ 2013, \apj, 772, 88 

\bibitem[Su \& van Ballegooijen(2012)]{Su2012}
Su, Y. \& van Ballegooijen, A., 2012, \apj, 757, 168

\bibitem[Tandberg-Hanssen \& Anzer(1970)]{Tandberg-Hanssen1970} Tandberg-Hanssen, E., \& Anzer, U.\ 1970, \solphys, 15, 158 

\bibitem[Tang(1987)]{Tang1987} Tang, F.\ 1987, \solphys, 107, 233 

\bibitem[Trujillo Bueno et al.(2002)]{TrujilloBueno2002}
Trujillo Bueno, J., Landi Degl'Innocenti, E., Collados, M., Merenda,
L., \& Manso Sainz, R. 2002, Nature, 415, 403

\bibitem[Trujillo Bueno \& Asensio Ramos(2007)]{TrujilloBueno2007} Trujillo Bueno, J., \& Asensio Ramos, A.\ 2007, \apj, 655, 642 

\bibitem[van Ballegooijen \& Martens(1989)]{vanBallegooijen1989} van Ballegooijen, A.~A., \& Martens, P.~C.~H.\ 1989, \apj, 343, 971 

\bibitem[van Ballegooijen \& Martens(1990)]{vanBallegooijen1990} van Ballegooijen, A.~A., \& Martens, P.~C.~H.\ 1990, \apj, 361, 283 

\bibitem[van Ballegooijen \& Su(2014)]{vanBallegooijen2014} van Ballegooijen, A.~A., \& Su, Y.\ 2014, Nature of Prominences and their Role in Space Weather, 300, 127 

\bibitem[Yeates \& Mackay(2009)]{Yeates2009} Yeates, A.~R., \& Mackay, D.~H.\ 2009, \solphys, 254, 77 

\bibitem[Yeates et al.(2007)]{Yeates2007} Yeates, A.~R., Mackay, D.~H., \& van Ballegooijen, A.~A.\ 2007, \solphys, 245, 87 

\bibitem[Zirker et al.(1997)]{Zirker1997} Zirker, J.~B., Martin, S.~F., Harvey, K., \& Gaizauskas, V.\ 1997, \solphys, 175, 27 

\end{thebibliography}
\end{document}